\documentclass[conference]{IEEEtran}
\IEEEoverridecommandlockouts
\usepackage{cite}
\usepackage{amsmath,amssymb,amsfonts}
\usepackage{algorithmic}
\usepackage{graphicx}
\usepackage{textcomp}
\usepackage{xcolor}

\usepackage{times}
\usepackage{soul}
\usepackage{url}
\usepackage[hidelinks]{hyperref}
\usepackage[utf8]{inputenc}
\usepackage[small]{caption}
\usepackage{graphicx}
\usepackage{amsmath}
\usepackage{amsthm}
\usepackage{booktabs}
\usepackage{algorithm}
\usepackage{algorithmic}
\usepackage{enumitem}
\usepackage{color}  
\usepackage{amsmath}
\usepackage{amssymb}
\usepackage{bm}
\usepackage{booktabs}
\usepackage{multirow}
\usepackage{graphicx}
\usepackage{subfigure}
\usepackage{hyperref}
\usepackage{algorithm}
\usepackage{algorithmic}
\usepackage{amsthm}
\usepackage[american]{babel} 
\newcommand\etal{\emph{et al.}}
\newcommand\ie{\emph{i.e.}}
\newcommand\eg{\emph{e.g.}}

\newcommand\etc{\emph{etc.}}
\usepackage[switch]{lineno}

\def\BibTeX{{\rm B\kern-.05em{\sc i\kern-.025em b}\kern-.08em
    T\kern-.1667em\lower.7ex\hbox{E}\kern-.125emX}}
\begin{document}



\title{Modeling Time-Variant User Interest for Click-Through Rate Prediction}


\title{Time Matters: A Novel Time-Aware  Long- and Short-term User Interest Model with Temporal Information for  Click-Through Rate Prediction} 

\title{Time Matters: A Novel Real-Time  Long- and Short-term User Interest Model for Click-Through Rate Prediction} 


\author{\IEEEauthorblockN{Xian-Jin Gui}
\IEEEauthorblockA{National Key Laboratory for Novel Software Technology\\
Nanjing University, Nanjing 210023, China\\
guixj@lamda.nju.edu.cn}

}

\maketitle

\begin{abstract}
Click-Through Rate (CTR) prediction is a core task in online personalization platform. A key step for CTR prediction is to learn accurate user representation to capture their interests. 
Generally, the interest expressed by a user is time-variant, \ie, a user activates different interests at different time. 
However, most previous CTR prediction methods overlook the correlation between the activated interest and the occurrence time, resulting in what they actually learn is the mixture of the interests expressed by the user at all time, rather than the real-time interest at the certain prediction time. 
To capture the correlation between the activated interest and the occurrence time, in this paper we investigate users’ interest evolution from the perspective of the whole time line and develop two regular patterns: periodic pattern and time-point pattern. Based on the two patterns, we propose a novel time-aware long- and short-term user interest modeling method to model users' dynamic interests at different time. 
Extensive experiments on public datasets as well as an industrial dataset verify the effectiveness of exploiting the two patterns and demonstrate the superiority of our proposed method compared with other state-of-the-art ones. 

\end{abstract}


\section{Introduction}
Click-Through Rate (CTR) prediction plays an important role in today's online personalization platform (\eg, e-commerce, online advertising, recommender systems), whose goal is to accurately predict  the probability of a user clicking a target item in certain context environments. 
Accurately modeling user interest is fundamental for CTR prediction task. 
In the past few years, some CTR prediction methods focusing on learning user interest have been proposed. DIN~\cite{zhou2018deep} is the pioneering work which pointed out that user's interests are diverse and proposed an attention-based mechanism to capture the relative interests to target item from user behaviors, but it ignores the temporal relation between behaviors. 
Later on, Zhou \etal~\cite{zhou2019deep} argued that user interests evolve over time dynamically and proposed a two-layer GRU model with attention mechanism to capture users' evolving interests. 
Along this line, Feng \etal~\cite{feng2019deep} observed that in some scenarios (\eg, e-commerce) users’ behavior sequences have the intrinsic structure that user behaviors are highly homogeneous in each session while heterogeneous cross sessions and proposed a method named DSIN to exploit the session information. Recently, Shi \etal~\cite{shi2020deep} proposed a time-stream framework which uses neural ODEs to integrate time interval information into user interest model. 
However, these studies capture user interest without considering  the correlation between the activated interest and the specific occurrence time, which causes what these methods actually model is the mixture of a user's diverse interests at all time, rather than the interest expressed by the user at the prediction time.  

\begin{figure}[t]
\centering
\includegraphics[width=\columnwidth]{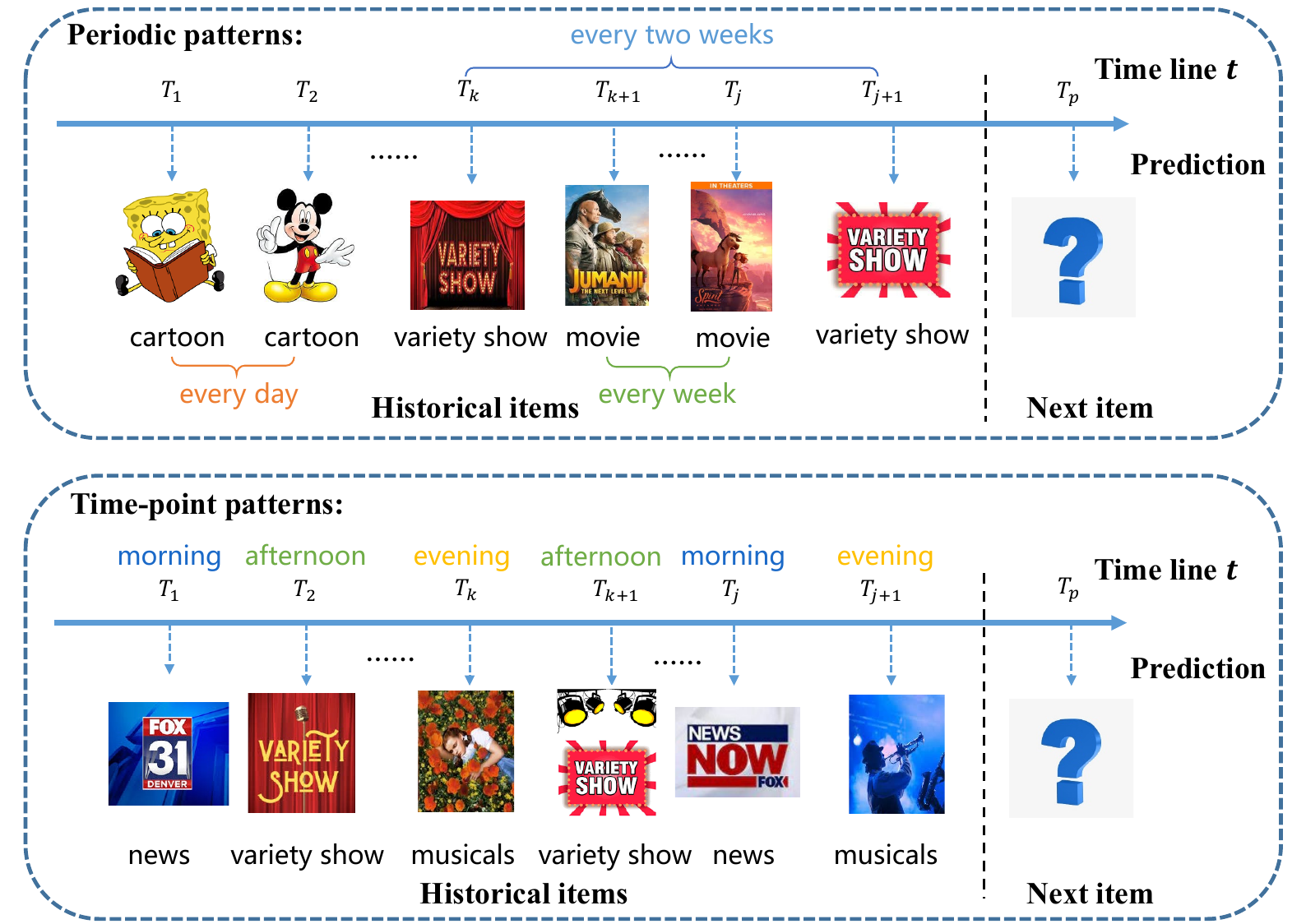}
\caption{An example of the periodic patterns and the time-point patterns.}
\label{fig:1}
\end{figure}

In our daily life, each person's behaviors usually exhibit some temporal regular patterns from the perspective of the whole time line.  As shown in Figure~\ref{fig:1},
let us consider a daily scenario: as a pastime, Mary watches a cartoon every day, a movie every week, and a variety show every two weeks. And she usually watches news videos in the morning, variety shows in the afternoon, and musicals in the evening on a video platform. Besides, as a heavy online shopper, she usually buys milk every month, daily snacks every week, \etc, and she usually buys school supplies at noon, daily necessities in the afternoon and some food at night on an e-commerce platform. 
Here we can see that the interest activated by a user is time-variant and the interest evolution usually exhibits two regular patterns: periodic pattern and time-point pattern. Periodic pattern means that the activation of a user's certain interest has certain time periods (\eg, every day or every week), while time-point pattern means that 
a user may express certain interest only at certain time points (\eg, morning or night). 
These two regular patterns usually exist simultaneously and interleave with each other. 
While the above examples seem a little artificial, in fact the similar patterns are common in our life. Some recent work focusing on  sequential recommendation tasks has also found that these patterns on Amazon datasets by case studies~\cite{ye2020time,li2020time}.

When capturing users’ interest evolution, it will be greatly beneficial to take the two patterns into consideration so as to capture the real-time interest a user activates at the prediction time. For example, when recommending items at night for a user, we need to focus on the interests that the user usually expresses at night rather than the mixture of the user's interests in the whole day. 
Since users have different occupations, spare times, tastes, lifestyles and so on, each user has her/his personalized temporal patterns of interest evolution. 
Further,  a user's interests are diverse, and the corresponding periodic pattern and time-point pattern may be various for different kinds of interests. Thus naively using rule-based methods with hard-coded periods and time points can not capture the personalized periodic pattern and time-point pattern of each user. Fortunately, each user's temporal patterns of different interests are hidden in her/his historical behaviors, and thus we can learn the temporal patterns from the user's behavior data. 

In this paper, we aim at modeling the real-time interest a user expresses on the target item at the prediction time. We develop the periodic pattern and time-point pattern, and propose a novel time-aware long- and short-term user interest modeling method by integrating them into model design. 
We summarize our main contributions as follows: 
\begin{itemize}
\item  We characterize periodic pattern and time-point pattern of users’ interest evolution to capture the correlation between the activated interest and the occurrence time.
\item We integrate the two patterns into the model design and propose a novel method to model the time-aware long- and short-term interests.
\item Extensive experiments on public datasets as well as an industrial dataset demonstrate the superiority of the proposed method compared with other state-of-the-art ones.
\end{itemize}


\section{Related Work} 
There are many existing methods related to CTR prediction, and we introduce the most related ones herein.

\subsection{Temporal Information}

The interest expressed by a user is time-variant, for example, it has been observed that the click behaviors of users over different news articles evolve over time in both Google News~\cite{liu2010personalized} and Yahoo News~\cite{zeng2016online}.
Since each user's behaviors happen on a time line, they naturally contain temporal information hidden in timestamps. 
Here are two kinds of temporal information that can be considered for CTR prediction, \ie, relative time interval information and absolute time point information. The relative time interval is important for capturing the association between users' behaviors, and further, the periodic pattern of user interest. 
while the absolute time point information can be used to capture the time-point pattern of user interest, since people may frequently interact with some specific items at certain time points. 
However, some early methods~\cite{cheng2016wide, paul2016deep, guo2017deepfm, zhou2018deep} ignore the temporal information of each behavior and treat each user's behaviors as a set, not considering any temporal relations between these behaviors. 
Later on, some work~\cite{hidasi2016session,zhou2019deep,feng2019deep} utilizes the ordinal relation of timestamps and take each user's behaviors as an ordered sequence for subsequently RNN-based modeling, but they do not exploit concrete temporal information conveyed in each behavior's timestamp. 
Recently, there are some studies noticing the importance of temporal information and devising some methods to exploit it. 
Zhu \etal~\cite{zhu2017what} and Yu \etal~\cite{yu2019adaptive} equipped LSTM structure with time-aware gates to capture temporal information, but they only considered time interval information, while ignoring time point information.  
Zhou \etal~\cite{zhou2018atrank} and Ye \etal~\cite{ye2020time} adopted time bucketization techniques to capture temporal information, while Wang \etal~\cite{wang2019regularized} extracted more fine-grained temporal information from timestamps and learned an embedding vector for each of them. And some other work incorporating temporal information into sequential recommendation tasks~\cite{ye2020time,li2020time,gligorijevic2019time,xu2019time}.
But these methods still do not explicitly capture time point information for modeling user interest.

\subsection{Long- and Short-term User Interest}


Generally, users' interests can be divided into long-term interests and short-term ones, which have different characteristics and are needed to model respectively. The \emph{long-term} interest is related to a user's personal tastes and is usually considered to be stable, which can be mined from the user's historical behaviors. The  \emph{short-term} interest tends to change frequently over time, and is reflected by a user's recent behaviors. Overall behaviors of a user may be determined by her/his long-term interests, but at any given time, a user is also affected by her/his short-term interests due to transient events such as new product releases. 
Previous studies~\cite{jannach2015adaptation} have pointed out that both of users'  long-term and short-term interests are of great importance for accurately capturing users' real-time interest. 

For CTR prediction, there are a few studies which consider modeling users’ long-term interests. Yu \etal~\cite{yu2019adaptive} adopted the attentive Asymmetric-SVD paradigm~\cite{koren2008factorization} to model users’ long-term interests. This method can only learn a static long-term interest representation, while overlooking that users' interests are multi-facet and vary when facing different target items. 
Actually, as pointed by Zhou \etal~\cite{zhou2018deep}, when given a target item, only part of interests will influence the action (to click or not), and thus we need to focus on the certain part of long-term interest of a user. 
Recently, rich historical behavior data of users are collected by online applications and they are proven to be of great value for better capturing users' interests~\cite{pi2019practice}. 
However, each user's historical behavior sequence contains lots of interactions corresponding to all kinds of interests. 
When tailored for a specific target item at some certain time, many historical behaviors in a user's behavior sequence may be irrelevant and even noisy to model current user interest on the target item. Thus, indiscriminately extracting user's interests from all historical behaviors will introduce a lot of irrelevant information, which may overwhelm effective signals. 

There are some studies trying to capture the temporal information of interest evolution, which mainly focus on  learning users' (short-term) interests from users' recent behaviors.
DIEN~\cite{zhou2019deep} adopts a two-layer GRU based model with attention mechanism to capture users' evolving interests, but it only leverages the ordinal information of behaviors. DSIN~\cite{feng2019deep} leverages session information of behavior sequence to model interest evolution. 
SLi-Rec~\cite{yu2019adaptive} uses a revised LSTM module with a time-aware controller capturing time interval information to model users’ short-term interests. 

Different from the above methods, we investigate users’ interest evolution from the whole time line and focus on the periodic pattern and time-point pattern.  
Especially, we simultaneously exploit absolute time point and relative time interval information and integrate them into long-term and short-term user interest model appropriately.



\section{Proposed Method} 
We first formulate the CTR prediction problem and introduce the notations. 
Let $\mathcal{U}=\{u_1, u_2, \dots, u_n\}$ denote the set of $n$ users and $\mathcal{V}=\{v_1, v_2, \dots, v_m\}$ denote the set of $m$ items. For a user $u$, she/he has a sequence of historical behaviors represented as $\mathcal{B}(u)=[(v_1^u, t_1^u), (v_2^u, t_2^u), \dots, (v_{|\mathcal{B}(u)|}^u, t_{|\mathcal{B}(u)|}^u)]$, where $(v_k^u, t_k^u)$ means user $u$ clicked item $v_k^u$ at timestamp $t_k^u\in\mathbb N$, $t_i^u < t_j^u$ for $i<j$ and $|\mathcal{B}(u)|$ denotes the number of interactions in the user's behavior sequence. The goal of CTR prediction is to predict the probability of user $u$ clicking a target item $v_p$ at a future prediction time $t_p$ (here $p = |\mathcal{B}(u)|+1)$, when the item $v_p$ is displayed to her/him. Following previous work, we take CTR prediction as a binary classification task.  
Given $\mathcal{B}(u)$, we will focus on user modeling to generate a dense user interest vector and take it together with other features as the input for binary classification. 

\subsection{Temporal Information Extraction}
Since users' behaviors happen on a time line, each timestamp $t_i^u\in\mathbb N$ actually contains concrete time point information: year, month, day, hour, minute, second, \etc ~This information further corresponds to the more meaningful time point information: morning/afternoon/night, weekday/weekend, seasons, \etc, which is valuable for capturing user time-variant interest. To capture the time point information, we construct a mapping function $\kappa(t)$: $\mathbb N \to \mathbb R^K$, which maps a timestamp $t$ to a $K$-dimension vector. Specifically, $\kappa(t) = [t.year, t.month$, $t.day, t.weekday, t.hour, t.minute, t.second]^T$. Note that $\kappa(t)$ implicitly encodes the time point information, such as seasons (by $t.month$), noon/afternoon (by $t.hour$), weekday/weekend (by $t.weekday$), \etc, which can capture the time-point pattern of interest evolution.
To capture the periodic pattern of interest evolution, 
we extract time interval information $(t_2-t_1)$ (assuming $t_1\le t_2$) for two behaviors happening at timestamps $t_1\in\mathbb N$ and $t_2\in\mathbb N$ respectively, which characterizes the length of time interval between the two behaviors and implicitly captures the period of some certain interest. 
To better characterize relative time interval, let $t_s$ denote the start time of the dataset (\ie, the smallest timestamp of behaviors in the dataset), 
we use $t - t_s$ to represent the relative time of each timestamp $t$.  In the following, we will use the temporal information $[t-t_s, \kappa(t)]\in\mathbb R^{K+1}$ of each behavior to capture \emph{periodic pattern} and \emph{time-point pattern} for users' long-term interests and short-term interests respectively.


\subsection{Modeling Long-term Interest}


Users' long-term interests are usually considered to be stable, 
which are reflected in users' long historical behavior data.  
Thus we can capture a user's long-term interests from her/his long-term behavior sequence. However, due to the diversity of user's interests and the uncertainty of user's behavior, given a target item, many behaviors in user's behavior sequence may be irrelevant or even noisy. Thus we need to focus on the certain part of the user’s long-term interest. To evade the influence of irrelevant behaviors, it is important to actively select relevant historical behaviors.
When tailored for a specific item, we propose to use content-based attention mechanism to obtain each historical item's attention score as follows:
\begin{align}
a_k^c&= \frac{\exp(x_kW^l_cx_p)}{\sum_{j=1}^{|\mathcal{B}(u)|}\exp(x_jW^l_cx_p)},
\label{eq:item-att}
\end{align}
where $x_k$ represents the embedding vector of the $k$-th item $v_k^u$ and $x_p$ represents the embedding vector of the target item $v_p$. 
Assume the dimension of embedding vector is $d$, then the dimension of the transformation matrix $W_c^l$ is $d\times d$.
The attention score $a_k^c$ determines which items should be emphasized or neglected according to the relevance between the contents of items and that of the target item.

Furthermore, the expression of user's long-term interests usually exhibits some certain temporal patterns, such as periodic pattern and time-point pattern,
since each person may express different interests at different time points. For example, when given a certain target item, although a user may be generally interested in it, but she/he may interact with it only on some special time points (\eg, morning or evening).  
This inspires us that the expression of a user's various interests may depend on the specific time point. Thus, when considering the interest that a user expresses at the prediction time, we need to take the prediction time $t_p$ into consideration. For capturing a user's interests at the prediction time $t_p$, we consider the periodic pattern and time-point pattern of the user interest with regard to $t_p$, both of which are hidden in users' behavior sequences. 
We design a novel temporal-based attention mechanism for integrating the temporal information to capture current user interest at $t_p$ as follows:
\begin{align}
a_k^t \!=\! \frac{\exp([t_k\!-\!t_s, \kappa(t_k)]W^l_t[t_p\!-\!t_s, \kappa(t_p)])}{\sum_{j=1}^{|\mathcal{B}(u)|}\exp([t_j\!-\!t_s, \kappa(t_j)]W^l_t[t_p\!-\!t_s, \kappa(t_p)])},
\label{eq:time-att}
\end{align}
where the dimension of the transformation matrix $W_t^l$ is $(K+1)\times (K+1)$.

Since the temporal-based attention mechanism is complementary to the content-based attention mechanism for capturing long-term interest on the target item $x_p$ at the prediction time $t_p$, we adopt concatenation operation to integrate these two modules. The final ``dynamic'' long-term interest representation is calculated as follows: 

\begin{equation}
p_u^{long} = \sum_{j\in B(u)}(a_j^c+a_j^t)x_j,
\end{equation}
which simultaneously combines the content-based mechanism and the temporal-based attention mechanism.
Note that different from previous methods, we learn a ``dynamic'' long-term interest representation, which focuses on the target item and the prediction time.

\subsection{Modeling Short-term Interest}
%
%

For users' short-term interests, it is important to capture users' time-varying states. 
Due to the remarkable ability of RNNs in sequential modeling, some work~\cite{zhou2019deep,yu2019adaptive} adopt them to model users' short-term interests. 
Among all RNN-based models, LSTM (long and short-term memory)~\cite{hochreiter1997long} and GRU (gated recurrent unit)~\cite{cho2014learning} are the most widely used for CTR prediction.  
Without loss of generality, in this paper we adopt LSTM as the base model and adapt it to capture users' time-variant short-term interests. But the same modification can also be applied to other RNN models (\eg, GRU). The formulation of vanilla LSTM is as follows: 
\begin{align}
   f_k   &= \sigma(W_fx_k+U_fh_{k-1}+b_f),\\
   i_k   &= \sigma(W_ix_k+U_ih_{k-1}+b_i),\\
\label{eq:c} c_k &=\! f_k \!\odot \!c_{k-1} \!+ \!i_k\!\odot\!\phi(W_cx_k\!+\!U_ch_{k-1}\!+\!b_c),\\
\label{eq:o}o_k  &= \sigma(W_ox_k+U_oh_{k-1}+b_o),\\
h_k  &= o_k\odot\phi(c_k),
\end{align}
where $W_*$, $U_*\in\mathbb{R}^{d\times d}$, $b_*\in\mathbb{R}^d$ are trainable parameters, $d$ denotes the dimension of input embedding and hidden state in RNN (we assume the dimensions are equal for notation clarity). $f_k$, $i_k$, and $o_k$ represent the forget, input, and output gates respectively. $c_k$ represents the cell status, $x_k$ represents the $k$-th item's embedding vector, and $\odot$ denotes the element-wise multiplication. while $\sigma$ represents the \emph{sigmoid} function and $\phi$ is the \emph{tanh} function.

Note that LSTM is originally designed to process words of sentence in NLP domain, where words can be regarded as evenly spaced and semantically consistent. But for sequential  behavior data, there are much complex relations. Firstly, time interval between two adjacent interactions can be various. Secondly, adjacent interactions in users' behavior sequences may not belong to the same semantic topic due to the mutability of users' intentions. 
Besides, as mentioned before, there are some regular patterns (periodic pattern and time-point pattern) hidden in users' behavior sequences that we need to consider for capturing a user's time-variant interest. Previous work~\cite{zhu2017what,yu2019adaptive} only considers time interval information 
between two behaviors' occurrence times $t_1$ and $t_2$ when modifying LSTM to model behavior sequence data, but overlooking the time point relevance between $t_1$ and $t_2$. 
Based on these considerations, we modify the gating logic of LSTM to better capture users' short-term interests. We firstly introduce a function which encodes the relation between $t_1$ and $t_2$:
\begin{equation}
\mathop{sim}(t_1, t_2) \!=\! h([t_1\!-\!t_s, \kappa(t_1)], [t_2\!-\!t_s, \kappa(t_2)]), \forall\, t_1, t_2,\\
\end{equation}
where $h(z_1, z_2) = \mathop{abs}(z_1-z_2)$, $\forall z_1$, $z_2\in\mathbb R^{K+1}$, with $\mathop{abs}(\cdot)$ denoting element-wise absolute value function. 
Here $\mathop{sim}(t_1,t_2)\in\mathbb{R}^{K+1}$ captures temporal distance (\emph{periodic pattern}) by ($t_1-t_s$, $t_2-t_s$) and time point relevance (\emph{time-point pattern}) by ($\kappa(t_1)$, $\kappa(t_2)$) simultaneously.
Then we introduce two time-aware features, \ie, adjacent time feature $\delta_{t_k}$ and time-span feature $s_{t_k}$ 
as follows:
\begin{align}
\delta_{t_k} &=\phi\left(W_{\delta}\mathop{sim}(t_{k-1},t_k)+b_\delta\right),\\
s_{t_k} &=\phi\left(W_{s}\mathop{sim}(t_{k},t_p)+b_s\right),
\end{align}  
where $W_\delta$, $W_s\in \mathbb R^{d\times {(K+1)}}$, $b_\delta, b_s \in \mathbb R^d$ are trainable parameters. The adjacent time feature $\delta_{t_k}$ encodes the temporal distance and the relevance of time points between two adjacent behaviors in behavior sequence, while the time-span feature $s_{t_k}$ encodes the temporal distance and the relevance between the $k$-th behavior's occurrence time $t_k$ and the prediction time $t_p$. 
To appropriately integrate the time-aware information into LSTM and adhere to the logic of LSTM, two gates are constructed as follows:
\begin{align}
T_\delta &= \sigma(W_{x\delta}x_k+W_{t\delta}\delta_{t_k}+b_{t\delta}),\\
T_s &= \sigma(W_{xs}x_k+W_{ts}s_{t_k}+b_{ts}),
\end{align}
where $W_{x\delta}$, $W_{xs}$, $W_{t\delta}$, $W_{ts}\in \mathbb R^{d \times d}$ and  $b_{t\delta}$, $b_{ts}$ $\in\mathbb R^d$.
Taking the two time-aware gates into consideration, we modify Eq.~\eqref{eq:c} to get
\begin{equation}
c_k \!=\! f_k \!\odot \! T_\delta\odot c_{k-1} + i_k\!\odot\!T_s\odot\phi(W_cx_k\!+\!U_ch_{k-1}\!+\!b_c),\\
\end{equation}
and modify Eq.~\eqref{eq:o} to get
\begin{align}
o_k  \!=\! \sigma(W_ox_k+ U_oh_{k-1} + W_{\delta o}\delta_{t_k}+W_{so}s_{t_k}+b_o).
\end{align}

The modified LSTM structure integrates the temporal information and implicitly models periodic pattern and time-point pattern, thus it can focus on the interest expressed by a user at prediction time $t_p$.
We can calculate all $h_j$ ($1$ $\le$ $j$$\le|\mathcal B(u)|$) with the modified LSTM structure.
Instead of directly using the last hidden state as the short-term interest representation, \ie, $p_u^{short} = h_{|\mathcal{B}(u)|}$, we adopt attention mechanism and formulate users' short-term interest representation as weighted average of all hidden states:
\begin{align}
a_k^s =& \frac{\exp(h_k W_h^s x_p)}{\sum_{j=1}^{|\mathcal{B}(u)|}\exp(h_j W_h^s x_p)},\\
p_u^{short} &= \sum_{j=1}^{|\mathcal{B}(u)|} a_j^sh_j,
\end{align}
where $x_p$ represents the embedding vector of the target item.

\begin{figure*}[t]
\centering
\includegraphics[width=2.0\columnwidth]{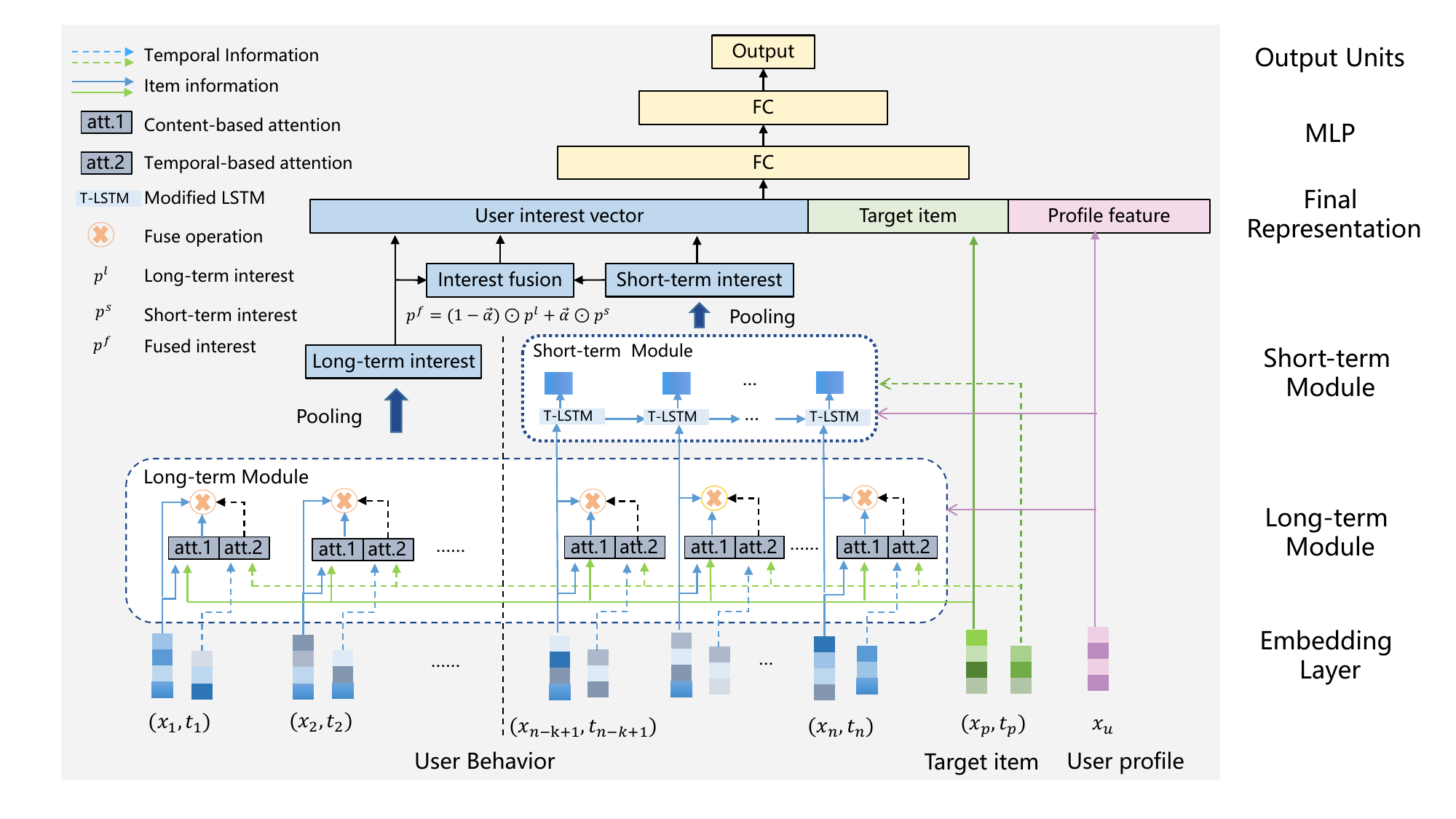} 
\caption{The overall structure of the TLSI model. The core part of the model is the long-term module and the short-term module, which generates long-term interest and short-term interest from a user’s behavioral sequence respectively.}
\label{fig:model}
\end{figure*}

\subsection{Long- and Short-term Interest Fusion} 

It is important to properly combine the user's long-term and short-term interests. 
Some previous studies~\cite{yu2019adaptive} directly linearly combine the long-term interest vector and the short-term one as the final interest vector of a user, but this linear scalar addition way may greatly limit the model capacity. In this paper, we propose to use a non-linear module $g$ to model the final user interest $p_u^{final}$ tailored for the target item $x_p$ and the prediction time $t_p$:
\begin{align}
p_u^{final} &= g\left(p_u^{long},\,p_u^{short}, \, u,\, x_{context}\right),
\end{align}
where $x_{context}$ is the context information that we need to consider, such as the target item $x_p$, the prediction time $t_p$, and the last behavior's occurrence time $ t_{|\mathcal{B}(u)|}$, \etc  ~When the long-term or the short-term interest is more relevant to the target item, it should obtain bigger weight than the other. When the prediction time $t_p$ is close to the last behavior's occurrence time $t_{|\mathcal B(u)|}$, the short-term interest may be more informative.
Inspired by Zhou \etal~\cite{zhou2019deep} and Lv \etal~\cite{lv2019sdm}, we design a gated neural network to fuse the long-term and short-term interest vectors.
A gate vector $\alpha_t^u\in{[0,1]}^{d}$ is calculated to adaptively decide the contribution percentages of long-term and short-term interests for the fused user interest on the target item $x_p$ at the prediction time $t_p$:  
\begin{align}
\alpha_t^u \!=\! \sigma&\left(W_g[p_u^{long}, \,p_u^{short}, x_{context}]\! + \!b_g\right),\\
\label{eq:fused} p_u^{fused} &=  \alpha_t^u\odot p_u^{long}+ (1-\alpha_t^u) \odot p_u^{short},
\end{align}
where the calculation of $\alpha_t^u$ dynamically relies on the long-term interest $p_u^{long}$, the short-term interest $p_u^{short}$ and the specific context information $x_{context}$.   

Instead of directly taking $p_u^{fused}$ as the final user interest vector, we concatenate $p_u^{long}$, $p_u^{short}$ and $p_u^{fused}$ as the final user interest vector:
\begin{align}
p_u^{final} &= \mathop{concat}\left([p_u^{long}, \,p_u^{short},\, p_u^{fused}]\right).
\end{align}
The rationale of concatenation is that only using $p_u^{fused}$ may lose valuable information in $p_u^{long}$ and $p_u^{short}$ when both of them are highly relevant to the prediction task and the information they contain is complementary. 
We name the overall model \textbf{TLSI} (\textbf{T}ime-aware \textbf{L}ong- and \textbf{S}hort-term user \textbf{I}nterest model).  
The overall model structure is shown in Figure~\ref{fig:model}.

\subsection{Training and Inference} 
Following the previous work~\cite{yu2019adaptive}, we concatenate the final user interest vector and the target item vector as the input of a two-layer MLP to focus on the impact of long-term and short-term interest module, \ie, $\hat{y} = \mathop{MLP}([p_u^{final},\,x_p])$, and all compared methods will share the same design. For CTR prediction, the negative log-likelihood function is widely used, which is defined as:
\begin{equation}
L = -\frac{1}{N}\sum_{i=1}^N\left(y_i\log \hat{y}_i + (1-y_i)\log(1-\hat{y}_i)\right)\label{eq:logloss},
\end{equation} 
where $N$ is the total number of training instances. And $y_i=1$ indicates a positive example (the user interacts with the item), while $y_i=0$ indicates a negative example. The optimization can be solved by minimizing the loss function with regularization techniques.

\section{Experiments}

We conduct experiments on the widely-used public datasets and an industrial dataset. The basic statistics of all these datasets are shown in Table~\ref{tab:dataset}. All source codes and the industrial dataset will be made publicly available to facilitate future studies.
\subsection{Evaluation Datasets}
\textbf{Amazon dataset\footnote{\url{http://jmcauley.ucsd.edu/data/amazon/}}}~\cite{mcauley2015image} is a dataset of user browsing logs from May 1999 to July 2014 over e-commerce products with reviews and product metadata collected from Amazon website.
We use four subsets of \emph{Amazon} dataset: \emph{CDs and Vinyl},  \emph{Movies and TV}, \emph{Electronics} and \emph{Books}. Following~\cite{zhou2018deep,zhou2019deep}, we regard all user reviews as user click behaviors. 
The timestamps of all \emph{Amazon} subsets are only accurate to the day and do not have more fine-grained hour information, which causes that these subsets are not very suitable for our model to exhibit superior performance. 

\textbf{Taobao dataset\footnote{\url{https://tianchi.aliyun.com/dataset/dataDetail?dataId=649&userId=1}}}~\cite{zhu2018learning} contains user behaviors collected from the commercial platform of Taobao.  
The dataset contains several types of user behaviors including click, purchase, add to shopping chart, favor, \etc\, The timestamps of these behaviors are accurate to the second (\ie, a resolution of seconds).
We take the click behaviors for each user and sort them according to the timestamps to construct users’ behavior sequences.


\textbf{The Industrial dataset} is constructed by collecting users' behavior logs and profile information from a mainstream video platform. Similar to public datasets, we collect features including user\_id, video\_id, category\_id, user watched video\_id \& category\_id lists and all behavior timestamps.  And the timestamps are also accurate to the second. In total, 539,980,407 samples are collected including 8,873,524 users, 2,399,239 videos, and 49,209 categories.  


According to the common experiment setting~\cite{zhou2019deep}, when there is no special declaration, the maximum length of each user's behavior sequence used for training on all datasets is set to be 100. 
Following previous studies~\cite{zhou2019deep,pi2019practice,ren19lifelong}, we use the first $|\mathcal B(u)|-1$ behaviors to predict whether the user $u$ will click the $|\mathcal{B}(u)|$-th item.

\begin{table}[tbp]
\centering
\caption{The statistics information of the datasets. }
\resizebox{0.48\textwidth}{!}{%
\begin{tabular}{@{}lccccc@{}}
\toprule
Dataset     & Users   & Items     & Categories & Interactions & Samples   \\ 
\midrule
 CDs          &  75,258   & 492,799  & 675    & 1,097,592     & 75,258       \\  
 Movies     & 123,960  & 208,321 & 345   & 1,697,533   & 123,960  \\ 
Electronics & 192,403  & 498,196  & 786  & 1,689,188  & 192,403   \\ 
Books       & 603,668  & 367,982   & 1,600    & 8,898,041 & 603,668   \\ 

\midrule
Taobao      & 987,994 & 4,162,024   &  9,439   & 100,150,807 & 987,994\\ 

Industrial      & 8,873,524   & 2,399,239   & 49,209   & 539,980,407  &   8,873,524 \\ 
\bottomrule
\end{tabular}%
}
\label{tab:dataset}
\end{table}

\subsection{Compared Methods}
We compare our proposed method TLSI with ten representative methods for CTR prediction task. 

\begin{itemize}

\item \textbf{ASVD}~\cite{koren2008factorization} directly represents users' (long-term) interest with the items that they have interacted with. All items appearing in a user's behavior sequence contribute equally.  
\item \textbf{A$^2$SVD}~\cite{yu2019adaptive} represents users' (long-term) interest by using weighted items, where the weighted scores are higher for behaviors which are more informative. However, the weighted scores of historical behaviors are invariant for different target items. 
\item \textbf{NARM}~\cite{li2017neural} is a neural attentive model, which captures a user's main purpose in the current session by incorporating an attention mechanism to RNN.  
\item \textbf{CA-RNN}~\cite{liu2016context} is the context-aware RNN model which employs adaptive context-specific input matrices and transition matrices in the RNN framework.
\item \textbf{DIN}~\cite{zhou2018deep} uses an attention mechanism to dynamically activate relevant items in a user's behavior sequence according to the target item. 

%
\item \textbf{LSTM}~\cite{hochreiter1997long} is a special form of RNN widely used in many NLP tasks, which can capture sequential relations.
%
\item \textbf{LSTM++}~\cite{yu2019adaptive} simultaneously use the A$^2$SVD and the LSTM module to capture both long-term and short-term interests respectively.  
\item \textbf{DIEN}~\cite{zhou2019deep} uses two-layer GRU with attention mechanism to capture users' evolving interests. It uses the calculated attentive values to control the second GRU layer and names it AUGRU. 
\item \textbf{Time-LSTM}~\cite{zhu2017what} adds time gates to model time intervals between two adjacent items in users' historical behavior sequence. Different from our method, it does not handle target item's contention and time point information. 
\item \textbf{SLi-Rec}~\cite{yu2019adaptive} uses collaborative filtering (CF) techniques to model users’ long-term interests and equips LSTM with time interval module to model user short-term interest. 
Different from our method, it does not consider time point information, and builds a static representation for users’ long-term interests without considering the specific target items and prediction time.  


\end{itemize}


We implement all the methods with Python 2.7 and Tensorflow 1.4, and the optimizer is the Adam. 
Dimension for item/category embedding
and RNN hidden layers is 18, while the dimension for RNN hidden state is 36. We adopt the Adam optimizer and the learning rate is set to 0.001. The number of training epochs is 10. The batch size is 128. We do not use dropout and the batch normalization is used only after the concatenation of the user’s (final) embedding and item embedding. The activation function of MLP is Dice~\cite{zhou2018deep}. The maximum length for user behaviors is set to 100. All experiments are repeated 3 times and the average results are reported. 


\textbf{Evaluation metric.} 
We take CTR prediction as a binary classification problem, and use Logloss in Eq.~\eqref{eq:logloss} and 
AUC in Eq.~\eqref{eq:auc} as the evaluation metric following previous studies~\cite{ouyang2019deep,ouyang2019representation,qin2020user}. 
Logloss measures the overall likelihood of the test data, and has been widely used for the classification tasks. 
AUC (Area Under ROC Curve) measures the probability that a positive example will be ranked higher than a randomly chosen negative one, which is defined as  
\begin{equation}
\label{eq:auc}
\text{AUC} = \frac{1}{m^+\cdot m^-} \sum_{x^+ \in D^+}\sum_{
x^-\in D^-}\mathbb I[f(x^+)>f(x^-)],
\end{equation}
where $f(\cdot)$ represents the CTR prediction model which gives the predicted click-through rate. $D^+$ and $D^-$ represent the set of positive examples and the set of negative examples respectively. The two metrics summarize a model’s performance from different aspects.


\subsection{Experiments on Public Datasets}  
\textbf{Comparison with other methods.} Table~\ref{tab:overall-auc} and Table~\ref{tab:log-loss} show the overall performances of different methods in terms of AUC and Logloss. ASVD, A$^2$SVD and DIN are the models without sequential module, but A$^2$SVD and DIN use attention mechanism. Comparing them, we can see that A$^2$SVD and DIN outperform ASVD, which justifies that assigning different relevance scores to items is beneficial.
The performance of LSTM++ is better than LSTM, which verifies that simultaneously exploiting users’ long-term and short-term interests is helpful. 
SLi-Rec achieves better performance than all other baselines, demonstrating that time interval information can help model user interest. At last, our method TLSI significantly outperforms all these methods including SLi-Rec on all five datasets. It is worth mentioning that we obtain the biggest performance gain on \emph{Taobao} dataset. This is due to that \emph{Taobao} dataset has sufficient training examples, and different from \emph{Amazon} dataset, each interaction in \emph{Taobao} dataset has accurate timestamp, both of which are beneficial to the capture of the periodic pattern and time-point pattern of user interest.


\begin{table}[tbp]
\centering
\caption{Performance comparison in terms of AUC. }
\resizebox{0.48\textwidth}{!}{%
\begin{tabular}{@{}lccccc@{}}
\toprule
Model       & Electronics              & Movies   &  CDs          & Books       & Taobao  \\ 
\midrule
ASVD            & 0.7250     &  0.7131 &  0.8115      &    0.7834   &  0.8113   \\
A$^2$SVD   &   0.7424   &  0.7182  &  0.8374     &   0.7991    &  0.8216     \\
DIN              &  0.7478    &  0.7208   &  0.8383     &  0.7974     &  0.8993    \\
\midrule
LSTM           &   0.7469   & 0.7273 &   0.8293      &    0.7915    &   0.8726     \\
LSTM++       &   0.7510   & 0.7302 &   0.8384      &    0.8007    &  0.8737       \\
NARM          &   0.7480   & 0.7249 &    0.8385     &    0.7990    &   0.8789     \\
CA-RNN       &  0.7551    & 0.7293 &  0.8282       &    0.7991    &   0.8660    \\
Time-LSTM  &   0.7541  & 0.7289 &   0.8254       &    0.8034    &  0.8719      \\
DIEN            &   0.7533   & 0.7277 &   0.8391     &   0.7982   &0.9017   \\
SLi-Rec       &   0.7659   &  0.7377      &  0.8394   &  0.8075  &  0.9049      \\
\midrule
TLSI           &   \textbf{0.7774}   &  \textbf{0.7437}      &    \textbf{0.8432}     &    \textbf{0.8115}  &   \textbf{0.9251}      \\
\bottomrule
\end{tabular}%
}
\label{tab:overall-auc}
\end{table}

{\color{blue}
\begin{table}[tbp]
\centering
\caption{Performance comparison in terms of Logloss. The smaller the value, the better the performance.}
\resizebox{0.48\textwidth}{!}{%
\begin{tabular}{@{}lcccccc@{}}
\toprule
Model       & Electronics              & Movies   &  CDs          & Books       & Taobao  \\ 
\midrule
ASVD            & 0.3253   &  0.3134  &  0.2690  &  0.3107   &  0.2645    \\
A$^2$SVD   & 0.3027    &  0.3151  &  0.2505   & 0.2988   &  0.2575     \\
DIN              & 0.2995    &  0.3133  &  0.2584   & 0.3113   &  0.1947    \\
\midrule
LSTM            &0.3002     &  0.3094  & 0.2550  &  0.3109    &   0.2199    \\
LSTM++       &0.2986   & 0.3083  &  0.2493   &  0.3005    &  0.2213     \\
NARM           & 0.3008    & 0.3109   & 0.2489    & 0.2984     &   0.2163   \\
CA-RNN       &  0.2970   &  0.3090  &  0.2551    &  0.2982    &  0.2244     \\
Time-LSTM  & 0.2969    & 0.3083   &   0.2571    &   0.2927    &   0.2225    \\
DIEN            &  0.2983   & 0.3096   & 0.2548  &  0.3004    & 0.1934    \\
SLi-Rec      &    0.2908  &  0.3064   &  0.2486  &  0.2943    &  0.1929     \\
\midrule
TLSI            &  \textbf{0.2848}   & \textbf{0.2998} &   \textbf{0.2476}  &  \textbf{0.2936}  & \textbf{0.1703}     \\
\bottomrule
\end{tabular}%
}
\label{tab:log-loss}
\end{table}
}

\textbf{Longer behavior sequences.} 
When capturing a user's long-term interests, it may be better to use her/his longer behavior sequence since more and more behavior data has been collected nowadays. 
To verify this, we set the maximum length of behavior sequence as 200 for the capture of long-term interest (we use ``longer'' to indicate this setting in Table~\ref{tab:longer}) and still set the maximum length as 100 for the capture of short-term interest. 
We test several methods with long-term interest module in experiments. The experimental results are shown in Table~\ref{tab:longer}. Comparing with Table~\ref{tab:overall-auc}, we can see that when using longer behavior sequences, all methods achieve better performance. This result inspires us that using longer behavior sequences to model long-term interest can further improve the performance of CTR prediction models.



\begin{table}[tbp]
\centering
\caption{Performance comparison in terms of AUC when using longer behavior sequences for long-term interest.}
\resizebox{0.48\textwidth}{!}{%
\begin{tabular}{@{}lcccccc@{}}
\toprule
Model       & Electronics             & Movies   &  CDs          & Books       & Taobao \\ 
\midrule
ASVD (longer)            & 0.7265     &  0.7143    &  0.8129      &    0.7842   & 0.8130    \\
A$^2$SVD (longer)   &   0.7435   &  0.7194    &  0.8382      &   0.8004     & 0.8226     \\
DIN (longer)              &  0.7482     &  0.7228    &  0.8396     &  0.7983     &  0.9013     \\
SLi-Rec (longer)        &   0.7673   &  0.7386      &  0.8405    &  0.8083      &  0.9064      \\
TLSI (longer)           &   \textbf{0.7786}   &  \textbf{0.7453}      &    \textbf{0.8457}     &    \textbf{0.8133}  &   \textbf{0.9262}    \\
\bottomrule
\end{tabular}%
}
\label{tab:longer}
\end{table}
\begin{table}[t]
\centering
\caption{Ablation study on temporal information (in AUC).}
\resizebox{0.48\textwidth}{!}{%
\begin{tabular}{@{}lcccccc@{}}
\toprule
Model       & Electronics & Movies &  CDs & Books & Taobao \\ 
\midrule
A$^2$SVD  &   0.7424   &  0.7182  &  0.8374     &   0.7991    &  0.8216   \\
DIN              &  0.7478    &  0.7208   &  0.8383     &  0.7974     &  0.8993   \\
DIEN            &   0.7453   & 0.7277 &   0.8303     &   0.7982   &0.9007   \\
SLi-Rec       &   0.7659   &  0.7377      &  0.8394   &  0.8075  &  0.9049    \\
TLSI-wo-TP     &   0.7755   &  0.7393      &  0.8413   &  0.8107 & 0.9172    \\
\midrule
A$^2$SVD+TP &   0.7432   &  0.7194  &  0.8381     &   0.8005    &  0.8234  \\
DIN+TP    &  0.7478    &  0.7208   &  0.8383     &  0.7974     &  0.8993   \\
DIEN+TP  &   0.7562   & 0.7288 &   0.8402     &   0.7997   &0.9028   \\
SLi-Rec+TP       &   0.7673   &  0.7387      &  0.8401   &  0.8082  &  0.9093      \\
TLSI           &   \textbf{0.7774}   &  \textbf{0.7437}      &    \textbf{0.8432}     &    \textbf{0.8115}  &   \textbf{0.9251}      \\
\bottomrule
\end{tabular}%
}
\label{tab:ablation-temporal}
\end{table}

\subsection{Ablation Study} 
Our method mainly comprises temporal information module, long-term interest module, short-term interest module and interest fusion module.
In the following, we investigate the effectiveness of them respectively. 

\textbf{The temporal information.} 
We test the effectiveness of temporal information and the model design of integrating them. Firstly, we want to verify whether our proposed time point information really contributes to performance improvement. Then, we want to check whether the performance gain only comes from the added temporal information. Thus we also need to verify that the model design also contributes to the performance improvement. 
To investigate the effectiveness of time point information, 
we construct A$^2$SVD+TP, DIN+TP, DIEN+TP and SLi-Rec+TP, which are hybrid models that incorporate time point information into base models by adding $\kappa(t)$ to model input. We also construct TLSI-wo-TP, the TLSI variant without utilizing time point information (we replace original $\kappa(t)$ values with meaningless all ones as placeholder). 
As shown in Table~\ref{tab:ablation-temporal},  the performance of TLSI-wo-TP is better than all compared methods. When adding time point information, the performance of all compared methods has been improved, but TLSI still outperforms all of them. The experimental results imply that both the time point information and the model design contribute to the performance improvement.

\begin{table}[tbp]
\centering
\caption{Comparison of variants of our model (in AUC).}
\resizebox{0.48\textwidth}{!}{%
\begin{tabular}{@{}lcccccc@{}}
\toprule
Model       & Electronics & Movies &  CDs & Books & Taobao \\ 
\midrule
A$^2$SVD   &   0.7424   &  0.7182  &  0.8374     &   0.7991    & 0.8216   \\
TLSI-L-c  & 0.7492     & 0.7233      &  0.8404  &  0.7998  & 0.9026        \\
TLSI-L-t  &  0.7497  &   0.7262     & 0.8334   &  0.7998 &0.8809           \\
TLSI-L       &  0.7564  & 0.7283       & 0.8406   &   0.8031  & 0.9188    \\
\midrule
LSTM           &   0.7469   & 0.7273 &   0.8293      &    0.7915    &   0.8726     \\
Time-LSTM  &   0.7541  & 0.7289 &   0.8254       &    0.8034    &  0.8719      \\
DIEN            &   0.7453   & 0.7277 &   0.8303     &   0.7982   &0.9007   \\
TLSI-S       &  0.7627    &0.7409      &   0.8287 &  0.8062  & 0.9044      \\
\midrule
TLSI-F &   0.7746   &  0.7413    & 0.8431   & 0.8093   &0.9216       \\
TLSI           &   \textbf{0.7774}   &  \textbf{0.7437}      &    \textbf{0.8432}     &    \textbf{0.8115}  &   \textbf{0.9251}      \\
\bottomrule
\end{tabular}%
}
\label{tab:only-long-term}
\end{table}

\textbf{Model variants.}
To investigate the effect of long-term interest, short-term interest and interest fusion, we delicately study several variants of TLSI. We use TLSI-L to represent the variant only with long-term interest (\ie, $p_u^{final}= p_u^{long}$), TLSI-S to represent the variant only with short-term interest (\ie, $p_u^{final}= p_u^{short}$), 
and TLSI-F denote the variant using the fused interest as the final interest (\ie, $p_u^{final}= p_u^{fused}$). Further, to verify whether our designs of content-based and temporal-based attention mechanisms are 
effective for long-term interest modeling, we use {TLSI-L-c} to represent the variant only with content-based attention mechanism (Eq.~\eqref{eq:item-att}) and TLSI-L-t to represent the variant only with temporal-based attention mechanism (Eq.~\eqref{eq:time-att}). We test all variants on all public datasets. Table~\ref{tab:only-long-term} shows the experimental results. All TLSI-L variants are better than A$^2$SVD in most cases, which demonstrates the superiority of considering dynamic long-term interest tailored for target item and the effectiveness of the two attention mechanisms. 
Note that TLSI-L is better than TLSI-S on \emph{CDs} and \emph{Taobao} datasets but inferior to TLSI-S on \emph{Electronics}, \emph{Movies} and \emph{Books} datasets, which implies both of them are indispensable and justifies the necessity of our design of using specially designed modules for them respectively. Comparing TLSI-F with TLSI-L and TLSI-S, we can see that the performance of TLSI-F is better than both of them, which verifies the effectiveness of fusing long-term and short-term interests (Eq.~\eqref{eq:fused}). But the performance of TLSI-F is inferior to TLSI, which justifies our conjecture that only using $p_u^{fused}$ as the final interest vector would lose some valuable information in $p_u^{long}$ and $p_u^{short}$.

\begin{figure}[t]  
\centering
\subfigure[different time intervals]{
\begin{minipage}[t]{0.473\linewidth}
\centering
\includegraphics[width=\linewidth]{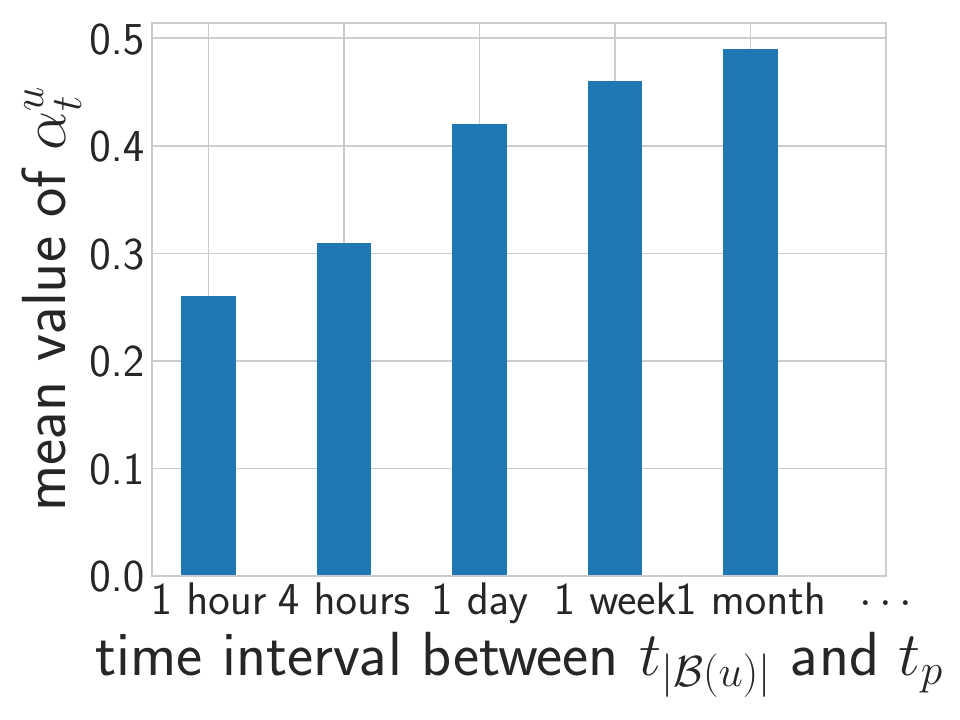}
\label{fig:gate-vector-a}
\end{minipage}%
}
\subfigure[different dimensions of  $\alpha_t^u$]{  
\begin{minipage}[t]{0.473\linewidth}
\centering
\includegraphics[width=\linewidth]{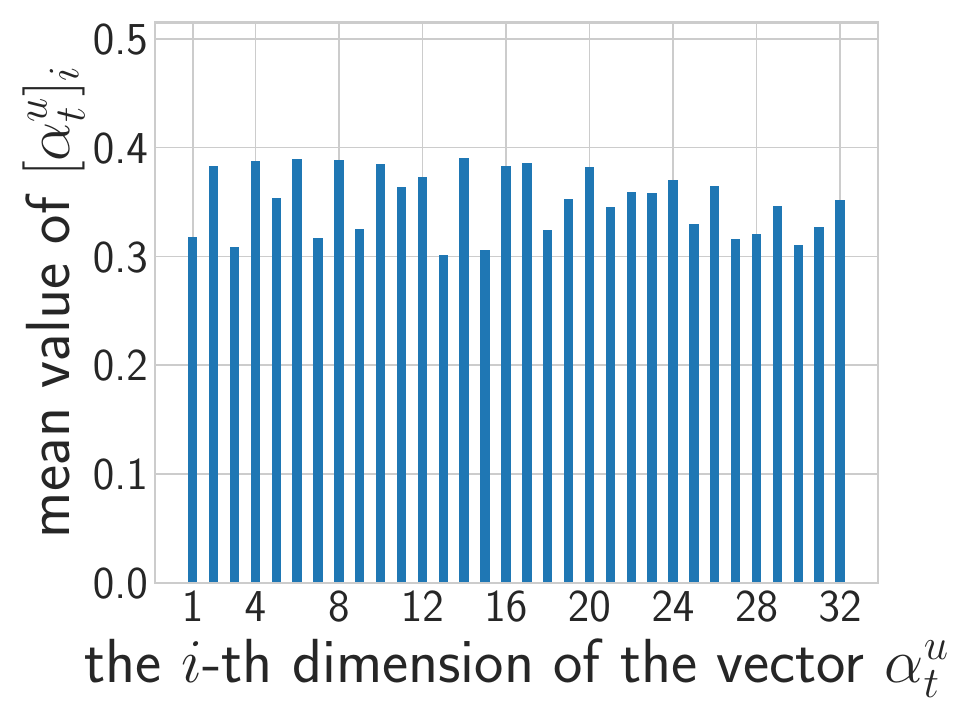} 
\label{fig:gate-vector-b}
\end{minipage}
}
\caption{The mean values of the gate vector  $\alpha_t^u$ for long-term and short-term interest fusion.}
\label{fig:gate-vector}
\end{figure}
\begin{figure}[t]  
\centering
\subfigure[content-based attention]{
\begin{minipage}[t]{0.473\linewidth}
\centering
\includegraphics[width=\linewidth]{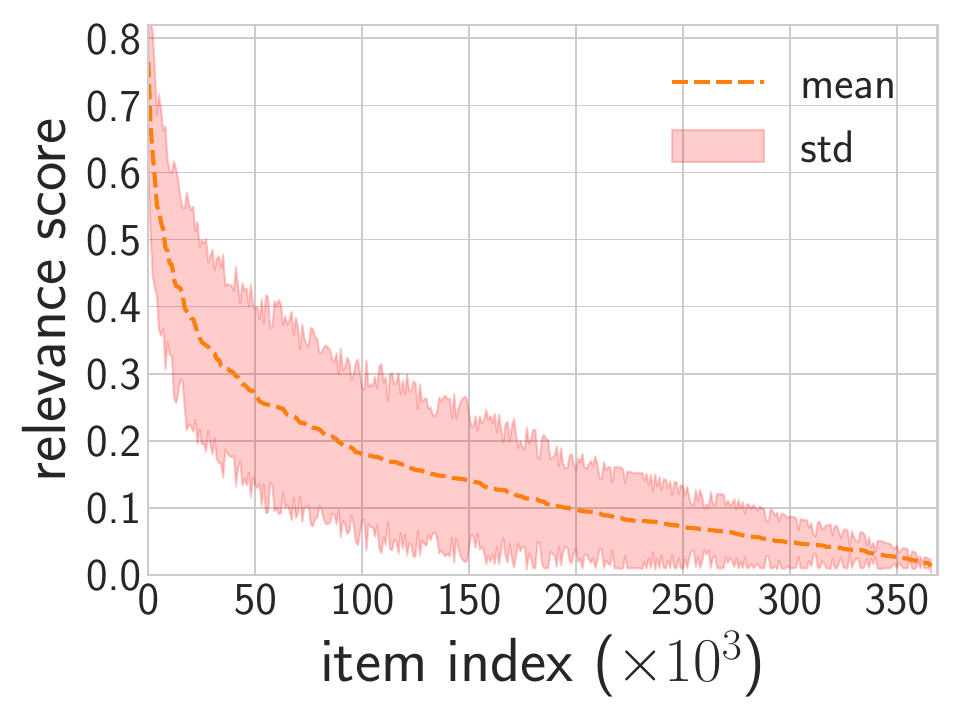}
\label{fig:att-long-a}
\end{minipage}%
}
\subfigure[temporal-based attention]{
\begin{minipage}[t]{0.473\linewidth}
\centering
\includegraphics[width=\linewidth]{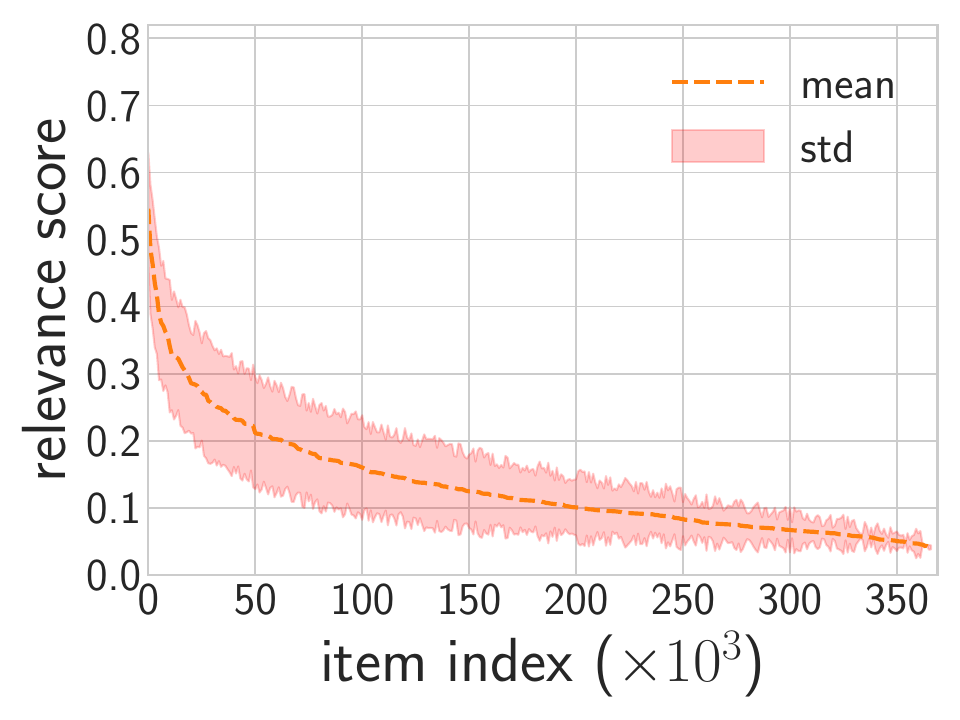} 
\label{fig:att-long-b}
\end{minipage}
}
\caption{Distribution of attentive weights (mean and std) of  two attention mechanisms for long-term interest module.}
\label{fig:att-long}
\end{figure}

\textbf{The interest fusion.}
To investigate the effect of the gate vector $\alpha_t^u$, we conduct two aspects of analysis on \emph{Books} dataset. 
Firstly, to get an intuitive sense of how the role of long-term interest changes with the context information, we split the test examples according to the value of the time interval between $t_p$ and $t_{|\mathcal B(u)|}$ (1 hour means $t_p-t_{|\mathcal B(u)|}\le$ 1 hour in Figure~\ref{fig:gate-vector-a}), and plot the mean values $\sum_{i=1}^d[\alpha_t^u]_i/d$ of the gate vector $\alpha_t^u$. 
As shown in Figure~\ref{fig:gate-vector-a}, there is an overall trend that the bigger the time interval between $t_{|\mathcal B(u)|}$ and $t_p$ is, the bigger weights the long-term interest gets
, which matches our previous conjecture. Then, to investigate  the effect of different dimensions of the gate vector $\alpha_t^u$, we plot the mean value of each dimension of  $\alpha_t^u$ in Figure~\ref{fig:gate-vector-b}. From Figure~\ref{fig:gate-vector-b}, we can see that the mean values of different dimensions of the gate vector $\alpha_t^u$ are different, which implies that using a gate vector rather than a scalar can capture the different importance of different dimensions of interest vectors.  

\textbf{Attention mechanisms for long-term interest.}  
To verify whether Eqs.~\eqref{eq:item-att} and~\eqref{eq:time-att} can learn discriminative relevance scores for historical behaviors, we calculate the mean value and the standard deviation of weighted values of each item on the \emph{Books} dataset for content-based and temporal-based attention mechanism respectively. 
We sort these items by mean values and plot the mean values and standard deviation.
As shown in Figure~\ref{fig:att-long}, different items are automatically assigned with different weights, which implies both content-based and temporal-based attention mechanisms can learn discriminative relevance scores. The standard deviation bars show that for different target items and prediction time, the same item will be assigned with different weights according to the specific context, since we capture the user's ``dynamic'' long-term interest.




\subsection{Experiments on the Industrial Dataset}
In the end, we test our method on the industrial dataset. 
As shown in Table~\ref{tab:indu}, our method achieves better performance than all compared methods. This result implies that our method has great practical application value for personalized recommendation, especially for video recommendation task. For the real video recommendation task, the correlation between the activated interest and the occurrence time is very important, and our method can capture ``dynamic'' user interest at the prediction time.


{\color{red}
\begin{table}[htbp]
\centering
\caption{Experimental results on Industrial dataset (in AUC and Logloss).}
\resizebox{0.48\textwidth}{!}{%
\begin{tabular}{@{}lcccccc@{}}
\toprule
Model       & ASVD & A$^2$SVD & DIN  & LSTM & LSTM++ & NARM\\ 
\midrule
AUC ($\uparrow$)   &   0.9234   &      0.9299    &     0.9471     &      0.9420    &     0.9431   &  0.9434\\
Logloss ($\downarrow$)   &   0.1765   &    0.1688     &    0.1487     &      0.1550    &     0.1540   &  0.1527 \\
\toprule
Model  & CA-RNN & Time-LSTM & DIEN  & SLi-Rec  &  TLSI & ---\\  
\midrule
AUC ($\uparrow$)     &    0.9388      &    0.9465     &   0.9533       &    0.9549   &   \textbf{0.9676} & ---\\
Logloss ($\downarrow$)     &    0.1585      &    0.1491     &   0.1375       &    0.1378   &   \textbf{0.1232} & ---\\
\bottomrule
\end{tabular}%
}
\label{tab:indu}
\end{table}
}

\section{Conclusion}




In this paper, we investigate interest evolution from the perspective of the whole time line and develop two regular patterns: periodic pattern and time-point pattern. The two regular patterns can well characterize the temporal patterns of interest evolution. Based on the two patterns, we propose a novel real-time long- and short-term user interest model for click-through rate prediction, which can model users' dynamic interests at different time. Extensive experiments on public datasets as well as an industrial dataset verify the effectiveness of exploiting the two patterns and demonstrate the superiority of our proposed method compared with other state-of-the-art ones. Especially, the experimental result on the industrial dataset implies that our method has great practical application value for personalized recommendation on video platforms.


\newpage


\bibliographystyle{IEEEtran}  
\bibliography{CTR-bib}  
\end{document}